\documentclass[final,3p,times]{elsarticle}
\usepackage{amsfonts}
\usepackage{amssymb}
\usepackage{amsmath}
\usepackage{graphics}
\usepackage{epsfig}
\usepackage{epstopdf}
\usepackage{amsthm}
\usepackage{textcomp}
\usepackage{color}
\usepackage{float}
\usepackage{multirow}
\usepackage{caption}
\biboptions{sort&compress}

\journal{Physica A}

\begin{document}

\begin{frontmatter}

\title{Maximizing the overall profit of a word-of-mouth marketing campaign: \\ A modeling study}

\cortext[cor1]{Corresponding author}

\author[label1]{Pengdeng Li}
\ead{1414797521@qq.com}

\author[label1]{Xiaofan Yang}
\ead{xfyang1964@gmail.com}

\author[label1,label2]{Lu-Xing Yang\corref{cor1}}
\ead{ylx910920@gmail.com}

\author[label1]{Qingyu Xiong}
\ead{Xiong03@equ.edu.cn}

\author[label1]{Yingbo Wu}
\ead{wyb@cqu.edu.cn}

\author[label3]{Yuan Yan Tang}
\ead{yytang@umac.mo}

\address[label1]{School of Software Engineering, Chongqing University, Chongqing, 400044, China}

\address[label2]{Faculty of Electrical Engineering, Mathematics and Computer Science, Delft University of Technology, Delft, GA 2600, The Netherlands}

\address[label3]{Department of Computer and Information Science, The University of Macau, Macau}

\begin{abstract}
As compared to the traditional advertising, the word-of-mouth (WOM) communications have striking advantages such as significantly lower cost and rapid delivery; this is especially the case with the popularity of online social networks. This paper addresses the issue of maximizing the overall profit of a WOM marketing campaign. A marketing process with both positive and negative WOM is modeled as a dynamical model knwn as the SIPNS model, and the profit maximization problem is modeled as a constrained optimization problem. The influence of different factors on the dynamics of the SIPNS model is revealed experimentally. Also, the impact of different factors on the expected overall profit of a WOM marketing campaign is uncovered experimentally.  On this basis, some promotion strategies are suggested. To our knowledge, this is the first time a WOM marketing campaign is treated this way.
\end{abstract}


\end{frontmatter}



\section{Introduction}

Promotion is a common form of product sales. The third-party advertising on mass media such as TV and newspaper has long been taken as the major means of promoting products. However, this promotion strategy suffers from high cost \cite{Armstrong2012, Grewal2016}. Furthermore, it has been found that, beyond the early stage of product promotion, the efficacy of advertising diminishes \cite{Goldenberg2001}.

Word-of-mouth (WOM) communications are a pervasive and intriguing phenomenon. It has been found that both satisfied and dissatisfied consumers tend to spread positive and negative WOM, respectively, regarding the products they have purchased and used \cite{Mahajan1984, Anderson1998}. As compared to positive WOM, negative WOM is more emotional and is more likely to influence the receiver's opinion. By contrast, positive WOM is more cognitive and more considered \cite{Herr1991, Charlett1995, Sweeney2007, Ahmad2014}. The significant role of WOM in product sales is supported by broad agreement among practitioners and academics. Both positive and negative WOM will affect consumers' purchase decisions.  Due to significantly lower cost and fast propagation, WOM-based promotion campaign outperforms their advertising counterparts \cite{Misner1999, Chevalier2006}. With the increasing popularity of online social networks such as Facebook, Myspace, and Twitter, WOM has come to be the main form of product marketing \cite{Trusov2009}. 

Currently, the major concern on WOM-based viral marketing focuses on the influence maximization problem: find a set of seeds such that the expected number of individuals activated from this seed set is maximized \cite{Peres2010}. Toward this direction, large number of seeding algorithms have been proposed \cite{Kempe2005, ChenW2009, Hinz2011, Dinh2014, Mochalova2014, Kempe2015, Samadi2016, ZhangHY2016a, Bharathi2007, ZhangHY2016b}. Additionally, a number of epidemic models capturing the WOM spreading process have been suggested \cite{Bass1969, Gardner2013, Sohn2013, LiS2013, LiS2014, Rodrigues2015, Rodrigues2016, JiangP2017, WangWD2003, YuY2003, WeiXT2013}. However, all of these works build on the premise that a single product or a few competing products are to be promoted.

Typically, a large-scale marketing campaign often involves a consistent collection of products, that is, a customer may purchase more than one product out of the collection. The ultimate goal of such a marketing campaign is to maximize the overall profit. To achieve the goal, it is crucial to determine those factors that have significant influence on the profit. To our knowledge, so far there is no literature in this aspect.

This paper addresses the issue of maximizing the overall profit of a WOM marketing campaign with rich products. A marketing process with both positive and negative WOM is modeled as a dynamical model known as the SIPNS model, and the profit maximization is modeled as an optimization problem. The influence of different factors on the dynamics of the SIPNS model is revealed experimentally. Also, the impact of different factors on the expected overall profit of a WOM marketing campaign is uncovered experimentally. On this basis, some promotion strategies are recommended. To our knowledge, this is the first time a WOM marketing campaign is treated this way.

The subsequent materials are organized as follows. Section 2 models a WOM marketing process as the SIPNS model, and models the profit maximization as an optimization problem. Section 3 experimentally reveals the influence of different factors on the dynamics of the SIPNS model. Section 4 experimentally uncovers the influence of different factors on the expected overall profit of a WOM marketing campaign. Finally, Section 5 summarizes this work and points out some future topics.

\newtheorem{rk}{Remark}
\newproof{pf}{Proof}
\newtheorem{thm}{Theorem}
\newtheorem{lm}{Lemma}
\newtheorem{exm}{Example}
\newtheorem{cor}{Corollary}
\newtheorem{de}{Definition}
\newtheorem{cl}{Claim}
\newtheorem{pro}{Proposition}
\newtheorem{con}{Conjecture}

\newproof{pfcl1}{Proof of Claim 1}
\newproof{pfcl2}{Proof of Claim 2}

\section{The problem and its modeling}

Suppose a marketer will launch a WOM promotion campaign for promoting a batch of products, with the goal of achieving the maximum possible profit. To achieve the goal, let us modeling the profit maximization problem as follows.

Let $[0, T]$ denote the time interval for the campaign, $M_t$ the target market at time $t$ for the campaign, $M(t) = |M_t|$ the number of individuals in $M_t$. Here, $M_t$ is classified as four categories, which are listed below.

\begin{enumerate}
  \item[(a)] \emph{Susceptible} individuals, i.e., those who haven't recently bought any product but are inclined to buy one.
  \item[(b)] \emph{Infected individuals}, i.e., those who have recently bought a product but haven't made comment on it.
  \item[(c)] \emph{Positive} individuals, i.e., those who have recently bought a product and have made a positive comment on it.
  \item[(d)] \emph{Negative} individuals, i.e., those who have recently bought a product and have made a negative comment on it.
 \end{enumerate}
 
 \noindent Let $S(t)$, $I(t)$, $P(t)$, and $N(t)$ denote the expected number at time $t$ of susceptible, infected, positive, and negative individuals in $M_t$, respectively. Then, $S(t)+I(t)+P(t)+N(t) = M(t)$.

Now, let us impose a set of statistical assumptions as follows.

\begin{enumerate}

    \item[(H$_1$)] At any time, the individuals outside of the target market enter the market at constant rate $\mu > 0$, where $\mu$ is referred to as the \emph{entrance rate}. 
    
    \item[(H$_2$)] At any time, a positive individual spontaneously exits from the market at constant rate $\delta_P$, where $\delta_P$ is referred to as the \emph{P-exit rate}.
    
    \item[(H$_3$)] At any time, an infected individual spontaneously exits from the market at constant rate $\delta_I$, where $\delta_I$ is referred to as the \emph{I-exit rate}.
    
    \item[(H$_4$)] At any time, a negative individual spontaneously exits from the market at constant rate $\delta_N$, where $\delta_N$ is referred to as the \emph{N-exit rate}.
    
    \item[(H$_5$)] Encouraged by positive comments, at time $t$ a susceptible individual buys a product (and hence becomes infected) at rate $\beta_P P(t)$, where the constant $\beta_P > 0$ is referred to as the \emph{P-infection force}.
    
    \item[(H$_6$)] Discouraged by negative comments, at time $t$ a susceptible individual exits from the market at rate $\beta_N N(t)$, where the constant $\beta_N > 0$ is referred to as the \emph{N-infection force}.

    \item[(H$_7$)] At any time, an infected individual makes a positive comment on the product he/she has recently bought (and hence becomes positive) at constant rate $\alpha_P > 0$, where $\alpha_P > 0$ is referred to as the \emph{P-comment rate}.
    
    \item[(H$_8$)] At any time, an infected individual makes a negative comment on the product he/she has recently bought (and hence becomes negative) at constant rate $\alpha_N > 0$, where $\alpha_N > 0$ is referred to as the \emph{N-comment rate}.
    
    \item[(H$_9$)] At any time, a positive individual is inclined to buy a new product (and hence becomes susceptible) at constant rate $\gamma_P > 0$, where $\gamma_P > 0$ is referred to as the \emph{P-viscosity rate}.
    
    \item[(H$_{10}$)] At any time, an infected individual is inclined to buy a new product (and hence becomes susceptible) at constant rate $\gamma_I > 0$, where $\gamma_I > 0$ is referred to as the \emph{I-viscosity rate}.
\end{enumerate}

Fig. 1 shows these assumptions schematically.

\begin{figure}[H]
	\setlength{\abovecaptionskip}{0.cm}
	\setlength{\belowcaptionskip}{-0.cm}
    \centering
    \includegraphics[width=0.6\textwidth]{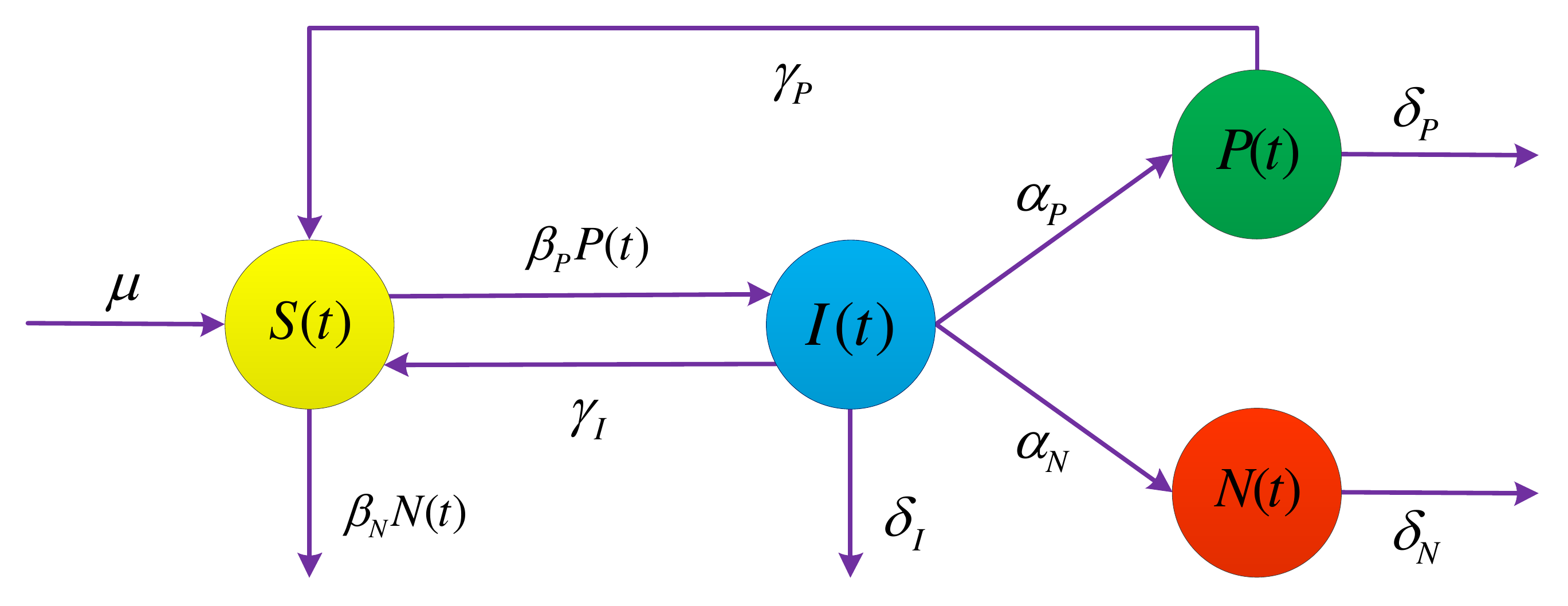}
    \caption{Diagram of assumptions (H$_1$)-(H$_{10}$).}
\end{figure}

This collection of assumptions can be formulated as the dynamical model
\begin{equation}
\left\{
\begin{aligned}
\frac{dS(t)}{dt}&=\mu -\beta_P P(t)S(t) - \beta_N N(t)S(t) + \gamma_PP(t) + \gamma_II(t),\\
\frac{dI(t)}{dt}&=\beta_P P(t)S(t) - \alpha_PI(t) - \alpha_NI(t) - \gamma_II(t) - \delta_I I(t),\\
\frac{dP(t)}{dt}&=\alpha_PI(t)-\gamma_PP(t) - \delta_P P(t),\\
\frac{dN(t)}{dt}&=\alpha_NI(t)-\delta_N N(t).\\
\end{aligned}
\right.
\end{equation}
Here, $0 \leq t \leq T$. We refer to the model as the Susceptible-Infected-Positive-Negative-Susceptible (SIPNS) model. 

Suppose the profit per item is one unit. Then the expected profit during the infinitesimal time interval $[t, t + dt)$ is $\beta_PP(t)S(t)dt$. As a result, the expected overall profit during the campaign is
\begin{equation}
    \beta _P \int_0^T P(t)S(t)dt.
\end{equation}
The marketer's goal is to maximize the expected overall profit subject to the SIPNS model.

\section{The dynamics of the SIPNS model}

The SIPNS model plays a key role in the overall profit of a WOM marketing campaign. However, it is difficult to theoretically study the dynamics of the four-dimensional SIPNS model. This section is intended to experimentally reveal the influence of different factors on the dynamics of the SIPNS model.

\subsection{The entrance rate}

Comprehensive simulation experiments show that, typically, the influence of the entrance rate on the dynamics of the SIPNS model is as shown in Fig. 2. In general, the following conclusions are drawn.

\begin{enumerate}
	\item[(a)] For any entrance rate, the expected number of susceptible, infected, positive and negative individuals levels off, respectively.
	\item[(b)] The steady expected number of susceptible individuals is not dependent upon the entrance rate.
	\item[(c)] With the rise of the entrance rate, the steady expected number of infected, positive, and negative individuals goes up, respectively. 
\end{enumerate}

\begin{figure}[ht]
	\setlength{\abovecaptionskip}{0.cm}
	\setlength{\belowcaptionskip}{-0.cm}
	\centering
	\includegraphics[width=0.6\textwidth]{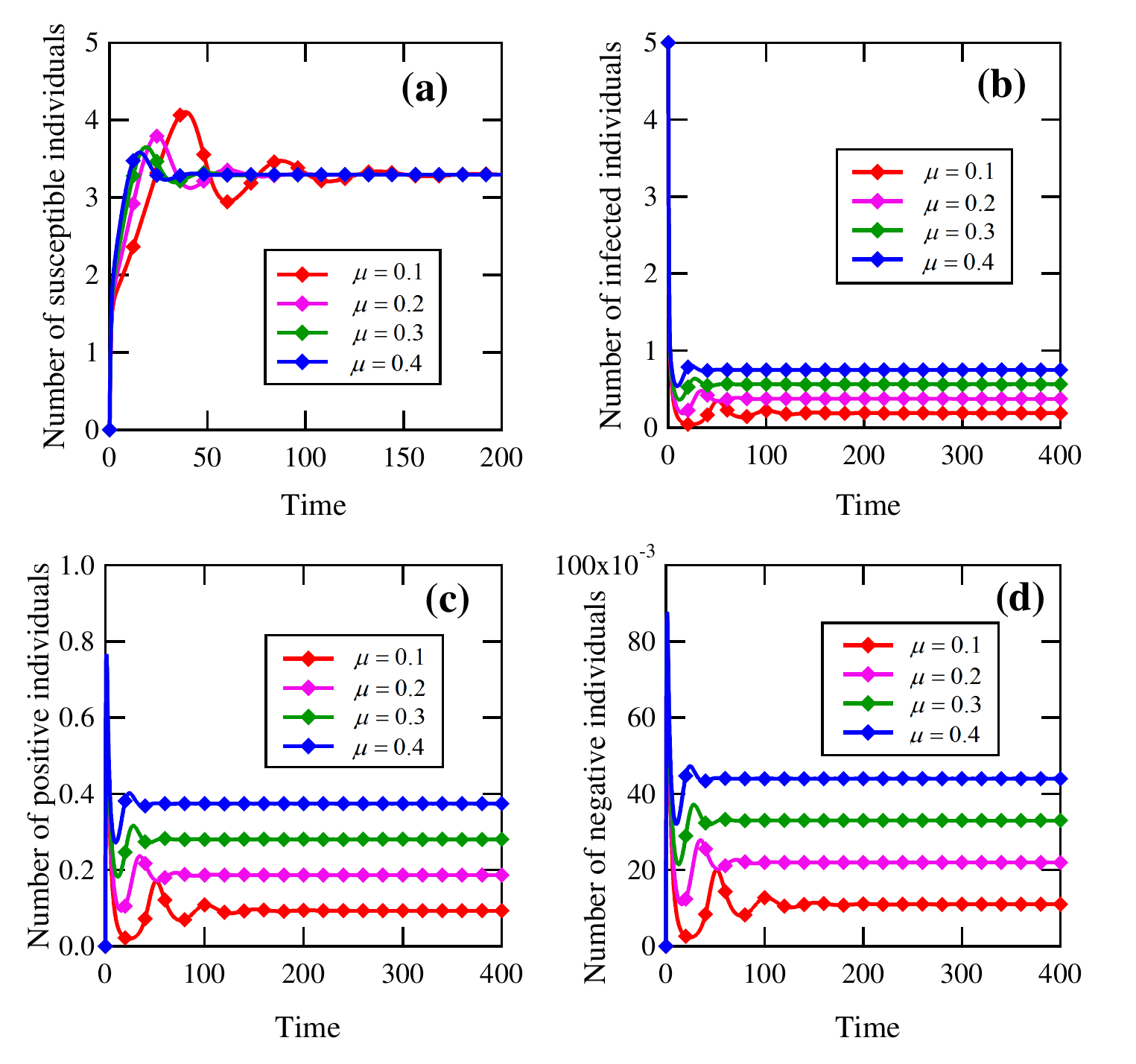}
	\caption{The time plots of $S(t)$, $I(t)$, $P(t)$, and $N(t)$ for different entrance rates.}
\end{figure}

\subsection{The I-exit rate}

Comprehensive simulation experiments show that, typically, the influence of the I-exit rate on the dynamics of the SIPNS model is as shown in Fig. 3. In general, the following conclusions are drawn.

\begin{enumerate}
	\item[(a)] For any I-exit rate, the expected number of susceptible, infected, positive, and negative individuals levels off, respectively.
	\item[(b)] With the rise of the I-exit rate, the steady expected number of susceptible individuals goes up. 
	\item[(c)] With the rise of the I-exit rate, the steady expected number of infected, positive, and negative individuals goes down, respectively.
\end{enumerate}

\begin{figure}[ht]
	\setlength{\abovecaptionskip}{0.cm}
	\setlength{\belowcaptionskip}{-0.cm}
	\centering
	\includegraphics[width=0.6\textwidth]{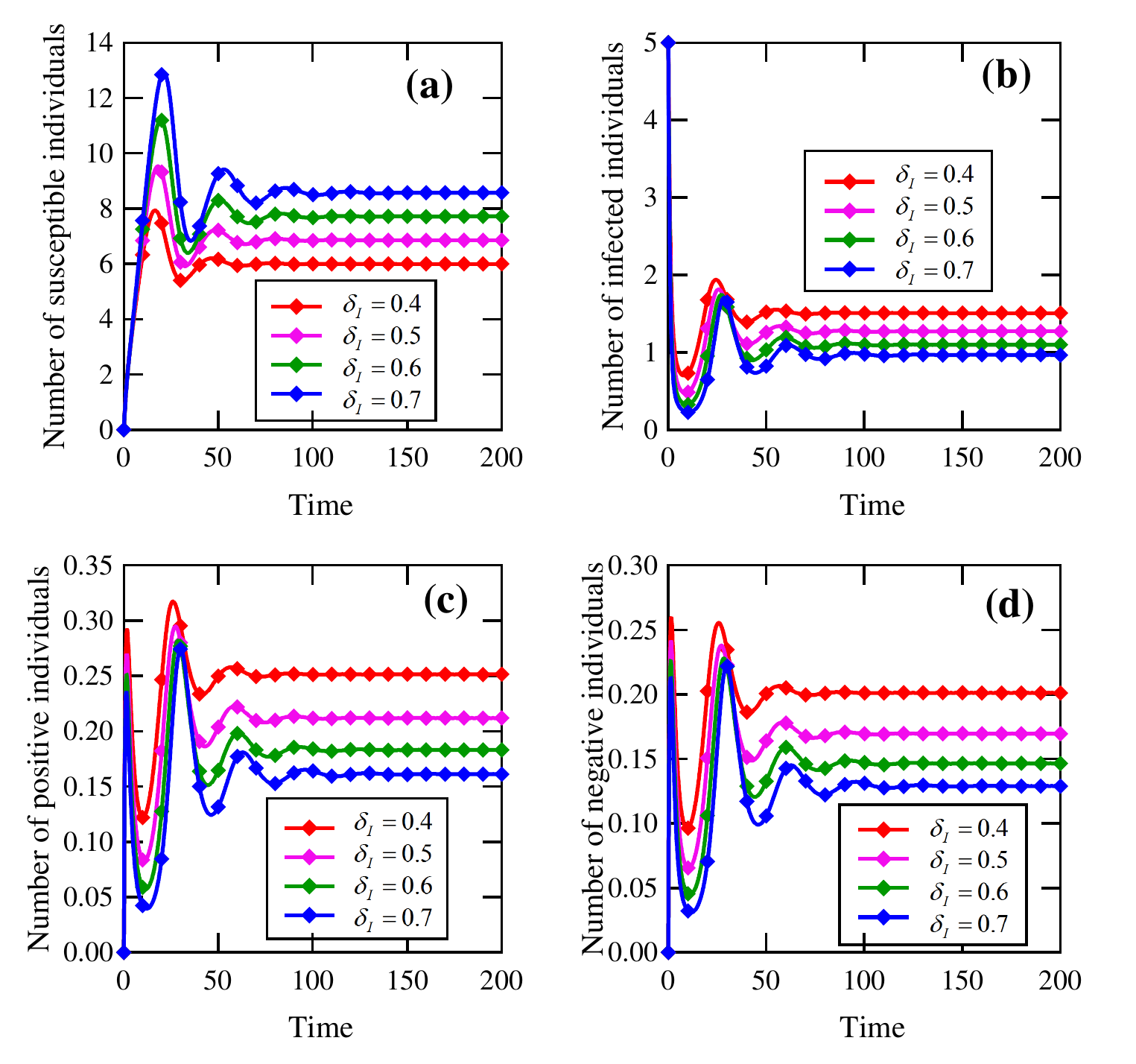}
	\caption{The time plots of $S(t)$, $I(t)$, $P(t)$, and $N(t)$ for different I-exit rates.}
\end{figure}

\subsection{The P-exit rate}

Comprehensive simulation experiments show that, typically, the influence of the P-exit rate on the dynamics of the SIPNS model is as shown in Fig. 4. In general, the following conclusions are drawn.

\begin{enumerate}
	\item[(a)] For any P-exit rate, the expected number of susceptible, infected, positive, and negative individuals levels off, respectively.
	\item[(b)] With the rise of the P-exit rate, the steady expected number of susceptible individuals goes up. 
	\item[(c)] With the rise of the P-exit rate, the steady expected number of infected, positive, and negative individuals goes down, respectively.
\end{enumerate}

\begin{figure}[ht]
	\setlength{\abovecaptionskip}{0.cm}
	\setlength{\belowcaptionskip}{-0.cm}
	\centering
	\includegraphics[width=0.6\textwidth]{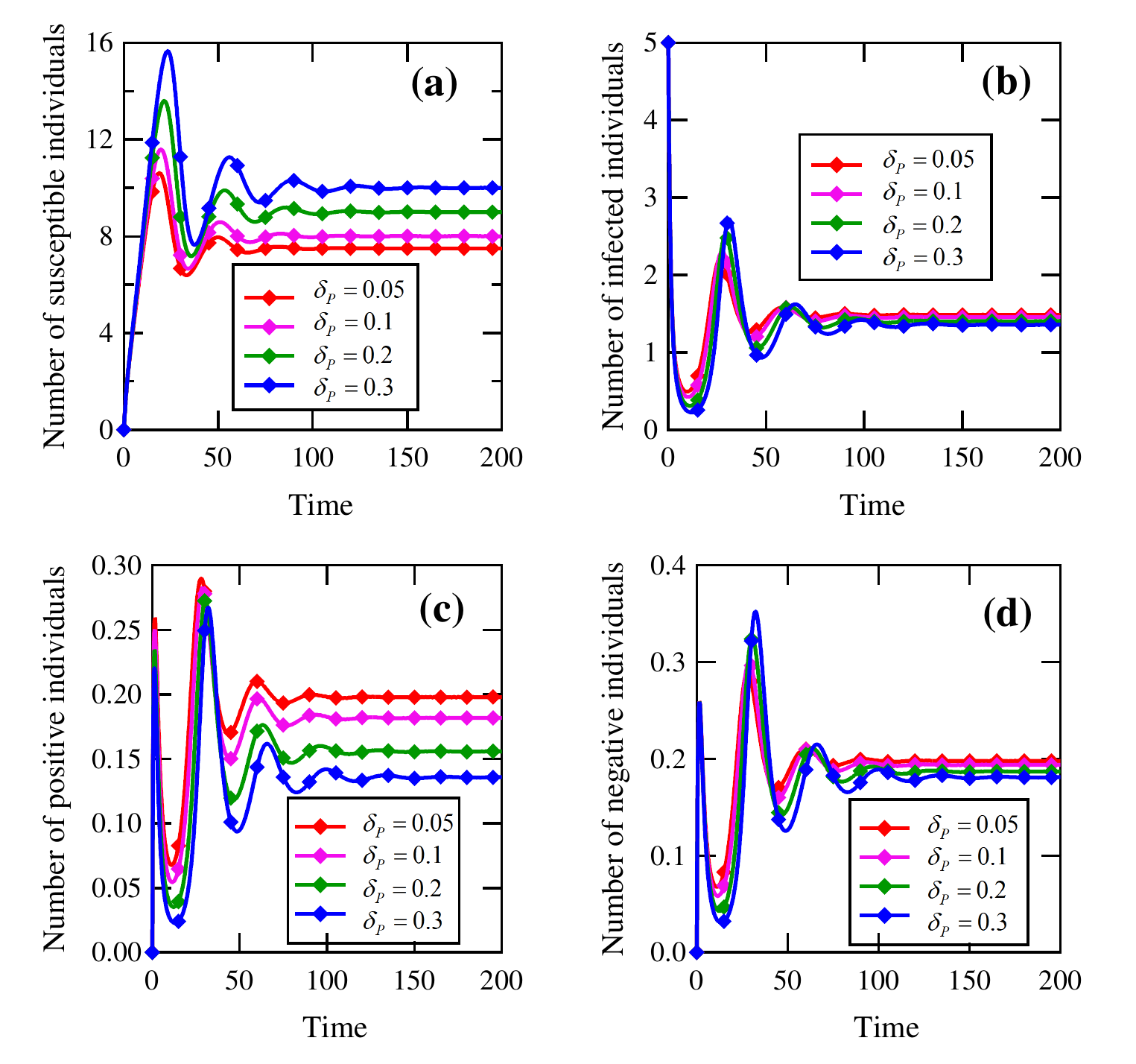}
	\caption{The time plots of $S(t)$, $I(t)$, $P(t)$, and $N(t)$ for different P-exit rates.}
\end{figure}

\subsection{The N-exit rate}

Comprehensive simulation experiments show that, typically, the influence of the N-exit rate on the dynamics of the SIPNS model is as shown in Fig. 5. In general, the following conclusions are drawn.

\begin{enumerate}
	\item[(a)] For any N-exit rate, the expected number of susceptible, infected, positive, and negative individuals levels off, respectively.
	\item[(b)] The steady expected number of susceptible individuals is not dependent upon the N-exit rate.
	\item[(c)] With the rise of the N-exit rate, the steady expected number of infected and positive individuals goes up, respectively.
	\item[(d)] With the rise of the N-exit rate, the steady expected number of infected and positive individuals goes down.
\end{enumerate}

\begin{figure}[ht]
	\setlength{\abovecaptionskip}{0.cm}
	\setlength{\belowcaptionskip}{-0.cm}
	\centering
	\includegraphics[width=0.6\textwidth]{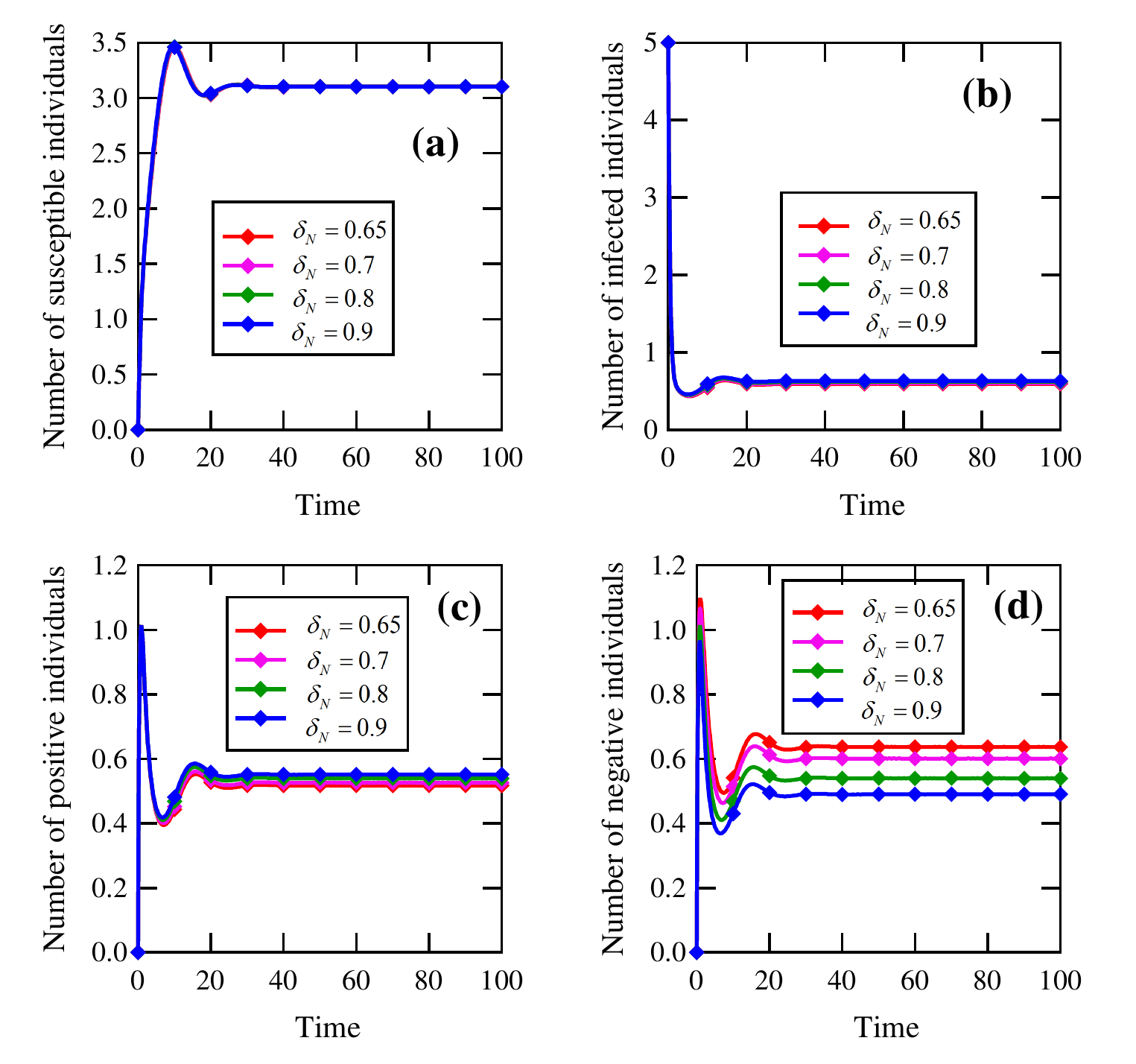}
	\caption{The time plots of $S(t)$, $I(t)$, $P(t)$, and $N(t)$ for different N-exit rates.}
\end{figure}

\subsection{The P-comment rate}

Comprehensive simulation experiments show that, typically, the influence of the P-comment rate on the dynamics of the SIPNS model is as shown in Fig. 6. In general, the following conclusions are drawn.

\begin{enumerate}
	\item[(a)] For any P-comment rate, the expected number of susceptible, infected, positive, and negative individuals levels off, respectively.
	\item[(b)] With the rise of the P-comment rate, the steady expected number of susceptible individuals goes down.
	\item[(c)] With the rise of the P-comment rate, the steady expected number of infected, positive, and negative individuals goes up, respectively.
\end{enumerate}

\begin{figure}[ht]
	\setlength{\abovecaptionskip}{0.cm}
	\setlength{\belowcaptionskip}{-0.cm}
	\centering
	\includegraphics[width=0.6\textwidth]{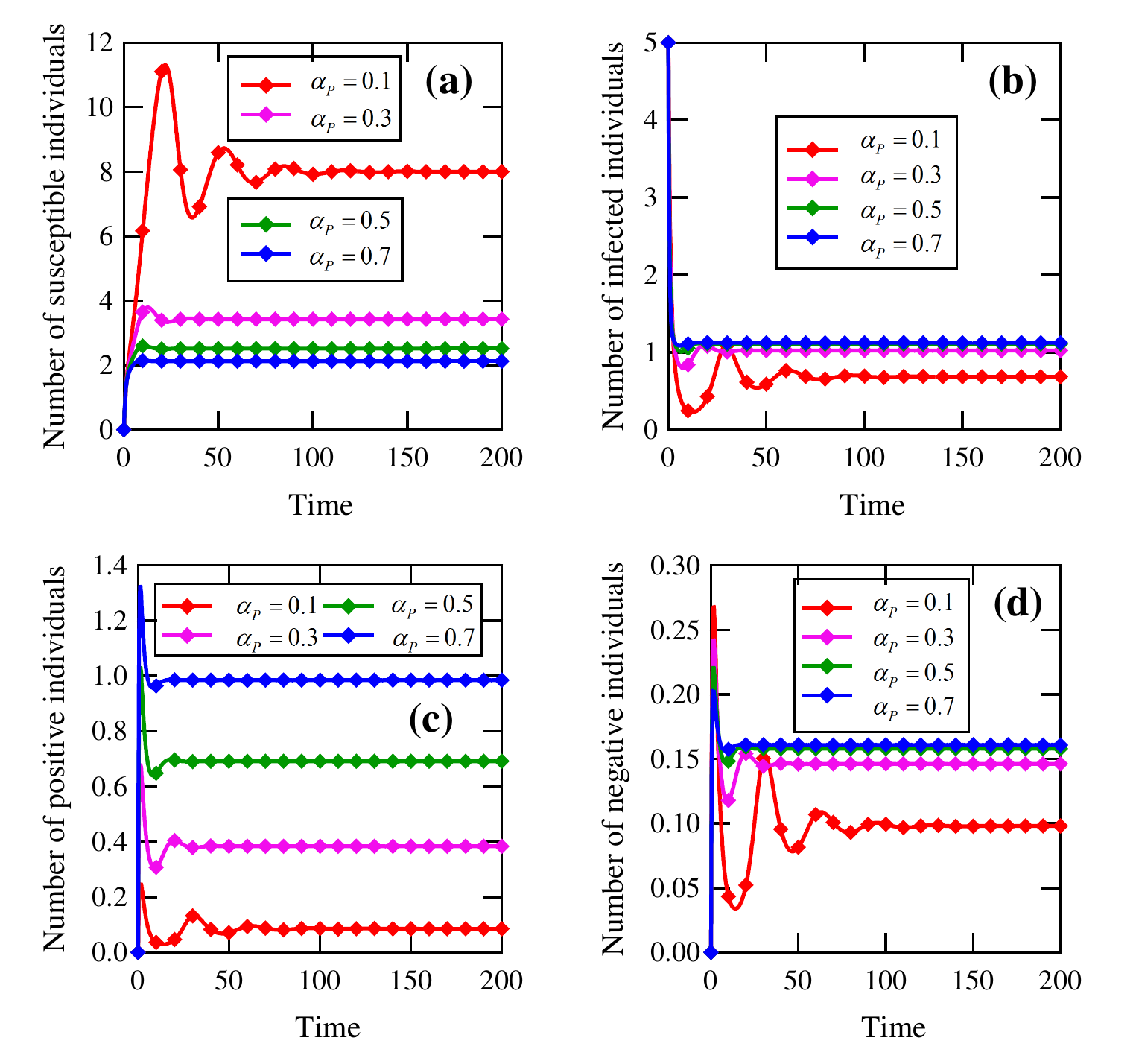}
	\caption{The time plots of $S(t)$, $I(t)$, $P(t)$, and $N(t)$ for different P-comment rates.}
\end{figure}

\subsection{The N-comment rate}

Comprehensive simulation experiments show that, typically, the influence of the N-comment rate on the dynamics of the SIPNS model is as shown in Fig. 7. In general, the following conclusions are drawn.

\begin{enumerate}
	\item[(a)] For any N-comment rate, the expected number of susceptible, infected, positive, and negative individuals levels off, respectively.
	\item[(b)] With the rise of the N-comment rate, the steady expected number of susceptible and negative individuals goes up, respectively.
	\item[(c)] With the rise of the N-comment rate, the steady expected number of infected and positive individuals goes down, respectively.
\end{enumerate}

\begin{figure}[ht]
	\setlength{\abovecaptionskip}{0.cm}
	\setlength{\belowcaptionskip}{-0.cm}
	\centering
	\includegraphics[width=0.6\textwidth]{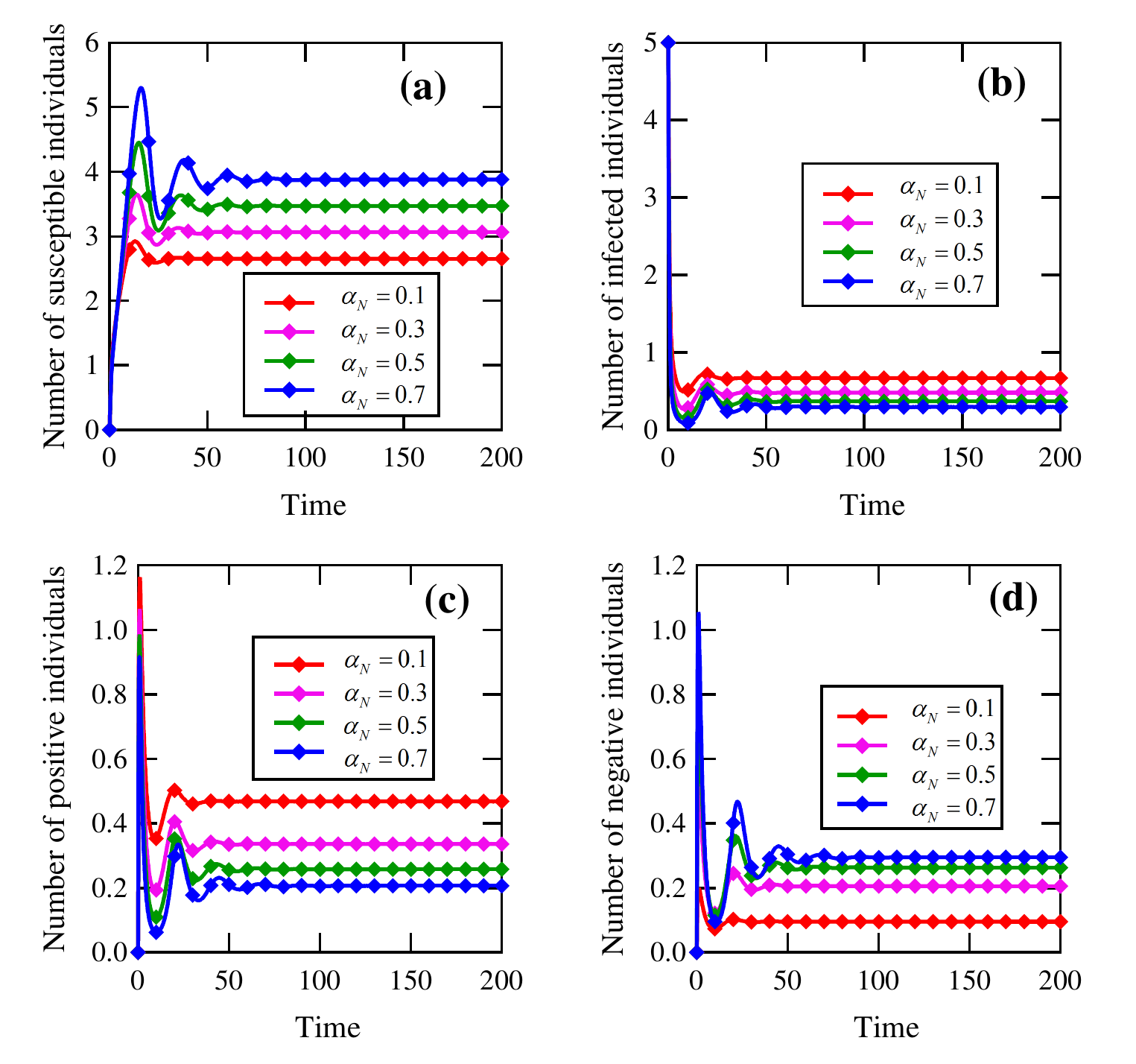}
	\caption{The time plots of $S(t)$, $I(t)$, $P(t)$ and $N(t)$ for different N-comment rates.}
\end{figure}

\subsection{The P-infection force}

Comprehensive simulation experiments show that, typically, the influence of the P-infection force on the dynamics of the SIPNS model is as shown in Fig. 8. In general, the following conclusions are drawn.

\begin{enumerate}
	\item[(a)] For any P-infection force, the expected number of susceptible, infected, positive, and negative individuals levels off, respectively.
	\item[(b)] With the rise of the P-infection force, the steady expected number of susceptible individuals goes down.
	\item[(c)] With the rise of the P-infection force, the steady expected number of infected, positive, and negative individuals goes up, respectively.
\end{enumerate}

\begin{figure}[ht]
	\setlength{\abovecaptionskip}{0.cm}
	\setlength{\belowcaptionskip}{-0.cm}
	\centering
	\includegraphics[width=0.6\textwidth]{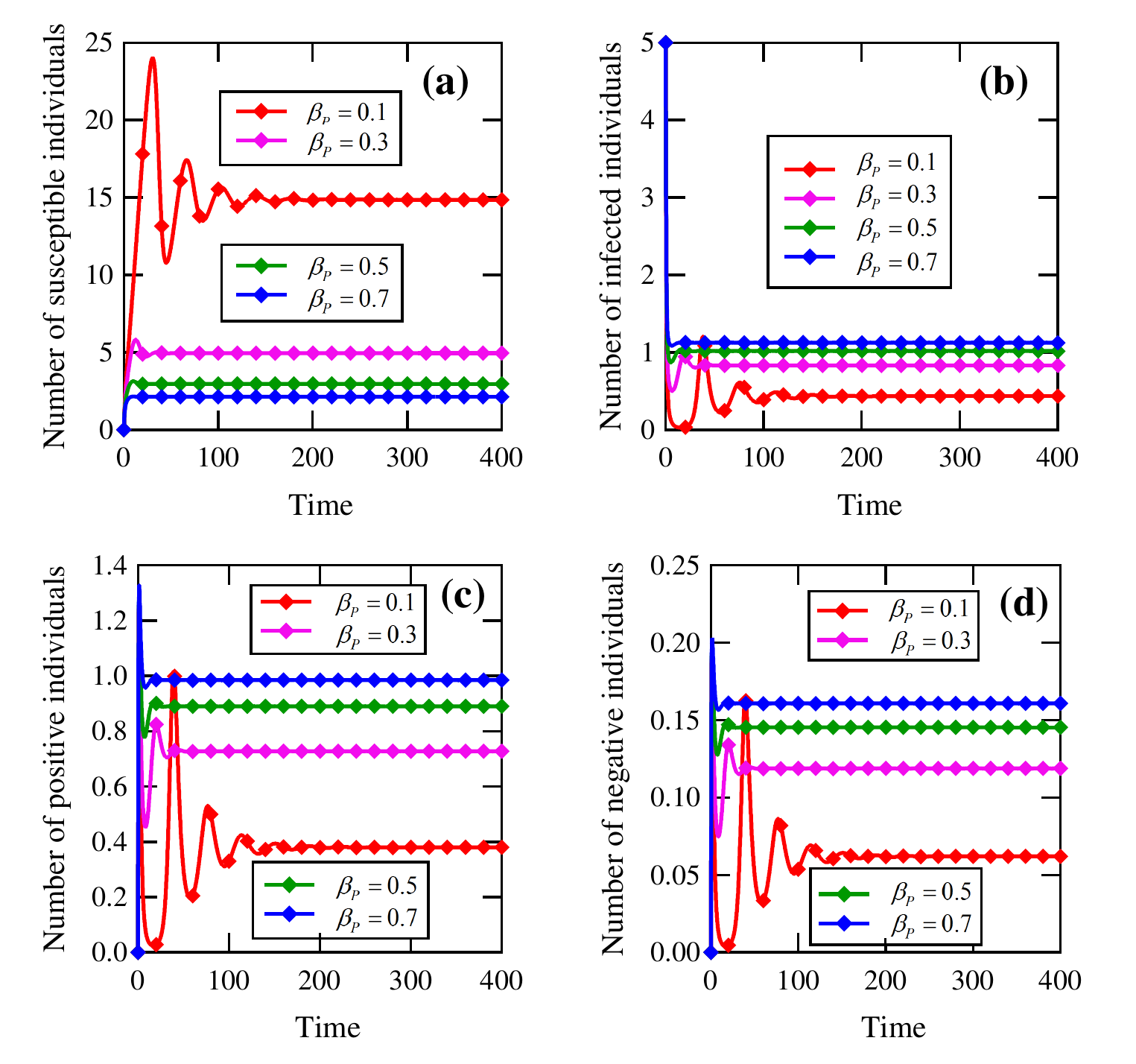}
	\caption{The time plots of $S(t)$, $I(t)$, $P(t)$ and $N(t)$ under different P-infection forces.}
\end{figure}

\subsection{The N-infection force}

Comprehensive simulation experiments show that, typically, the influence of the N-infection force on the dynamics of the SIPNS model is as shown in Fig. 9. In general, the following conclusions are drawn.

\begin{enumerate}
	\item[(a)] For any N-infection force, the expected number of susceptible, infected, positive, and negative individuals levels off, respectively.
	\item[(b)] The steady expected number of susceptible individuals is not dependent upon the N-infection force.
	\item[(c)] With the rise of the N-infection force, the steady expected number of infected, positive, and negative individuals goes down, respectively.
\end{enumerate}

\begin{figure}[ht]
	\setlength{\abovecaptionskip}{0.cm}
	\setlength{\belowcaptionskip}{-0.cm}
	\centering
	\includegraphics[width=0.6\textwidth]{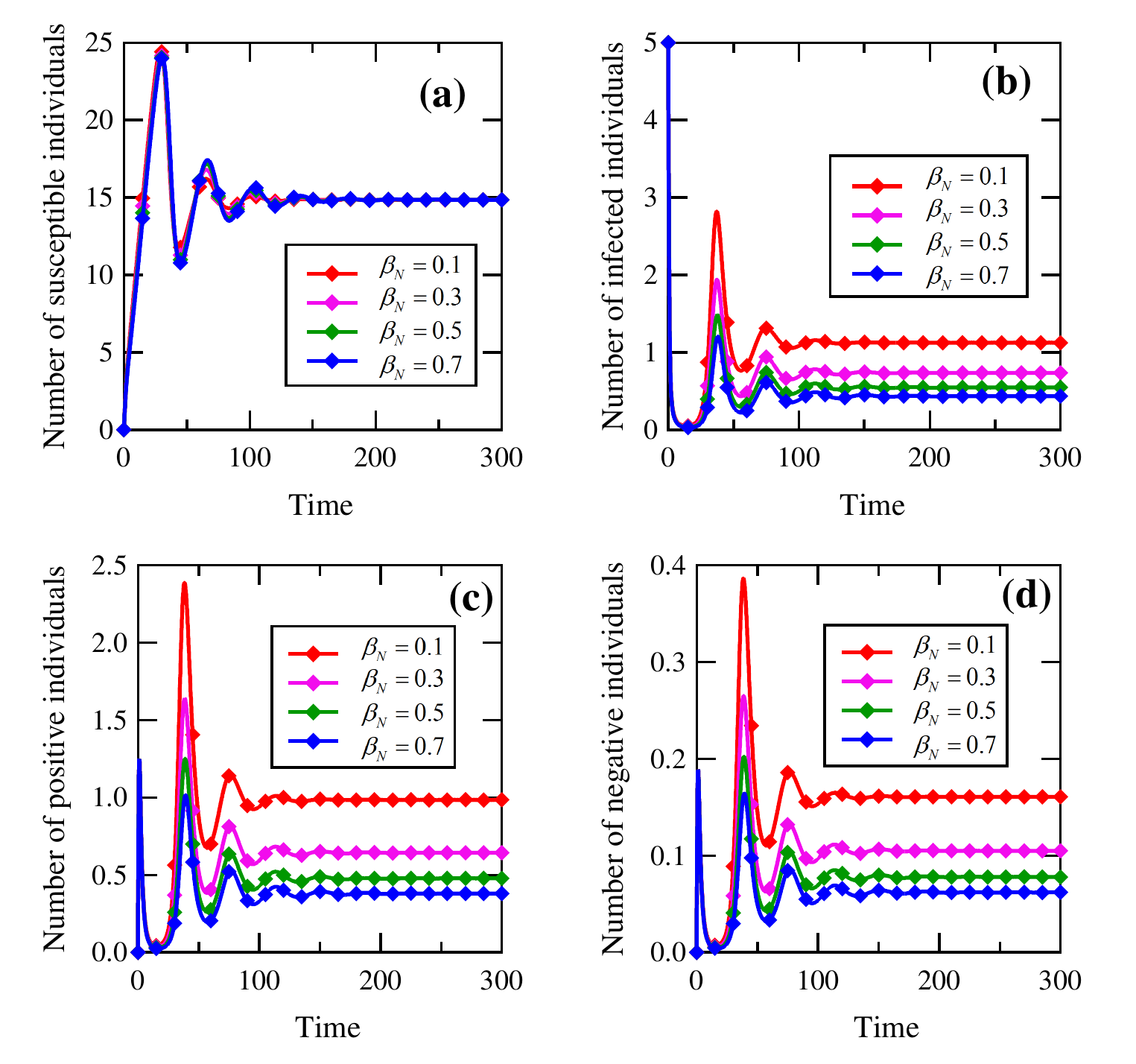}
	\caption{The time plots of $S(t)$, $I(t)$, $P(t)$ and $N(t)$ under different N-infection forces.}
\end{figure}

\subsection{The I-viscosity rate}

Comprehensive simulation experiments show that, typically, the influence of the I-viscosity rate on the dynamics of the SIPNS model is as shown in Fig. 10. In general, the following conclusions are drawn.

\begin{enumerate}
	\item[(a)] For any I-viscosity rate, the expected number of susceptible, infected, positive, and negative individuals levels off, respectively.
	\item[(b)] With the rise of I-viscosity rate, the steady expected number of susceptible individuals goes up.
	\item[(c)] With the rise of the I-viscosity rate, the steady expected number of infected, positive, and negative individuals goes down, respectively.
\end{enumerate}

\begin{figure}[ht]
	\setlength{\abovecaptionskip}{0.cm}
	\setlength{\belowcaptionskip}{-0.cm}
	\centering
	\includegraphics[width=0.6\textwidth]{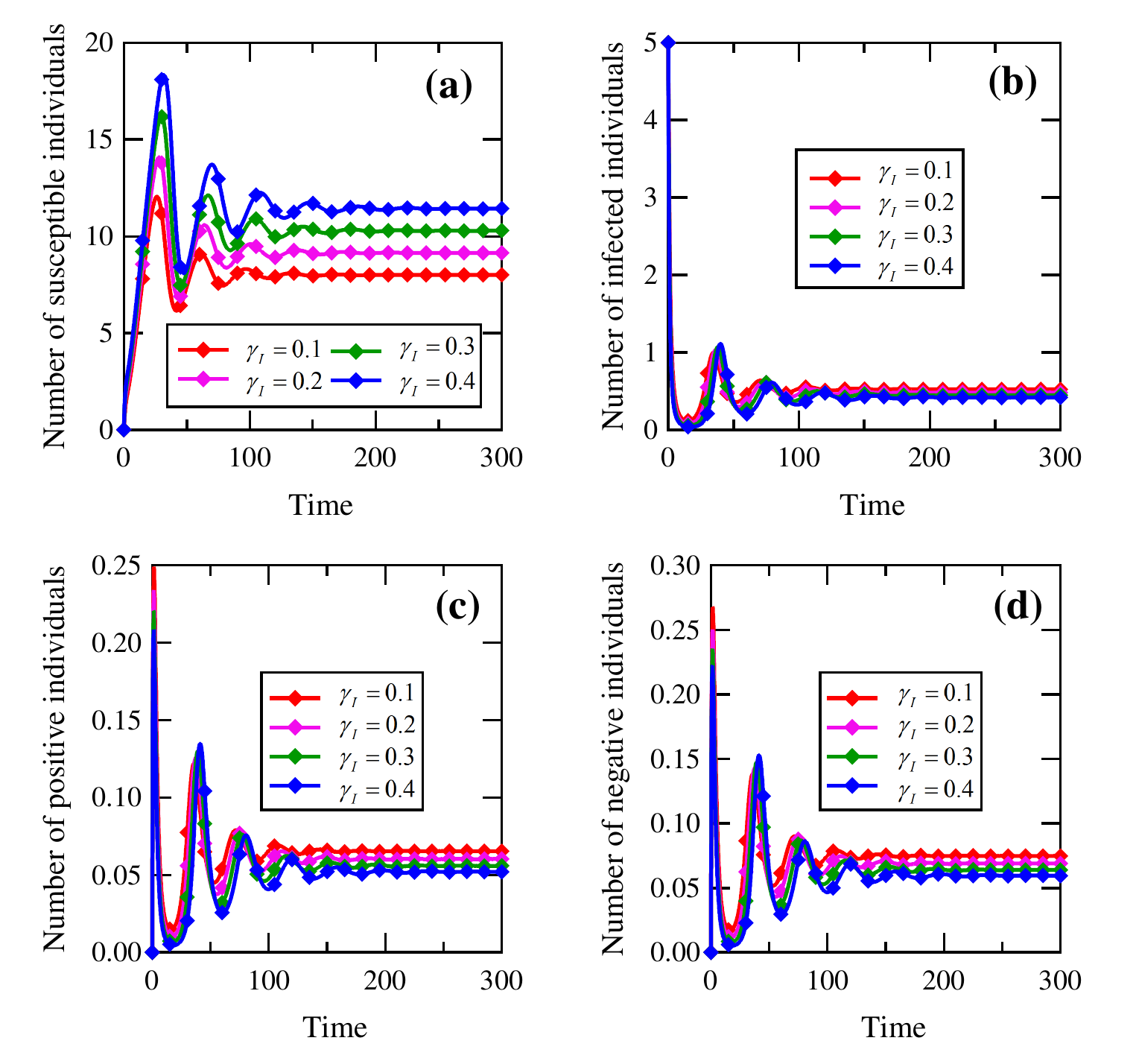}
	\caption{The time plots of $S(t)$, $I(t)$, $P(t)$, and $N(t)$ for different I-viscosity rates.}
\end{figure}

\subsection{The P-viscosity rate}

Comprehensive simulation experiments show that, typically, the influence of the P-viscosity rate on the dynamics of the SIPNS model is as shown in Fig. 11. In general, the following conclusions are drawn.

\begin{enumerate}
	\item[(a)] For any P-viscosity rate, the expected number of susceptible, infected, positive, and negative individuals levels off, respectively.
	\item[(b)] With the rise of P-viscosity rate, the steady expected number of susceptible individuals goes up.
	\item[(c)] With the rise of the P-viscosity rate, the steady expected number of infected, positive, and negative individuals goes down, respectively.
\end{enumerate}

\begin{figure}[ht]
	\setlength{\abovecaptionskip}{0.cm}
	\setlength{\belowcaptionskip}{-0.cm}
	\centering
	\includegraphics[width=0.6\textwidth]{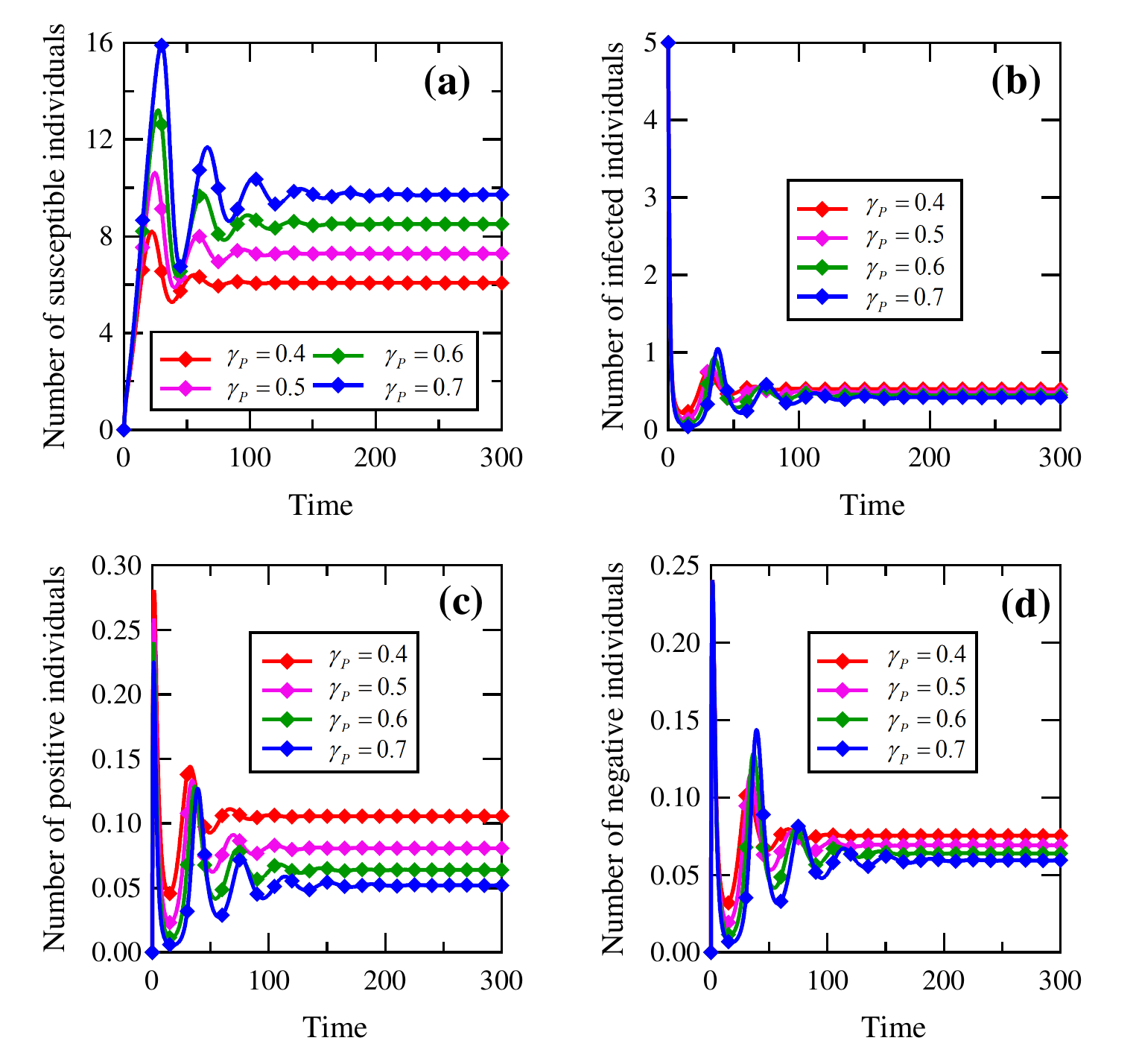}
	\caption{The time plots of $S(t)$, $I(t)$, $P(t)$, and $N(t)$ for different P-viscosity rates.}
\end{figure}

\section{The overall profit of a WOM marketing campaign}

This section aims to experimentally uncover the influence of different factors on the expected overall profit of a WOM marketing campaign.

\subsection{The entrance rate and the exit rates}

Comprehensive simulation experiments show that, typically, the influence of the entrance rate and the three exit rates on the expected overall profit is as shown in Fig. 12. In general, the following conclusions are drawn.

\begin{enumerate}
	\item[(a)] With the rise of the entrance rate, the expected overall profit goes up.
	\item[(b)] With the rise of the I-exit rate entrance rate, the expected overall profit goes down.
	\item[(c)] With the rise of the P-exit rate entrance rate, the expected overall profit goes down.
	\item[(d)] With the rise of the N-exit rate entrance rate, the expected overall profit goes up.
\end{enumerate}

\begin{figure}[ht]
	\setlength{\abovecaptionskip}{0.cm}
	\setlength{\belowcaptionskip}{-0.cm}
	\centering
	\includegraphics[width=0.5\textwidth]{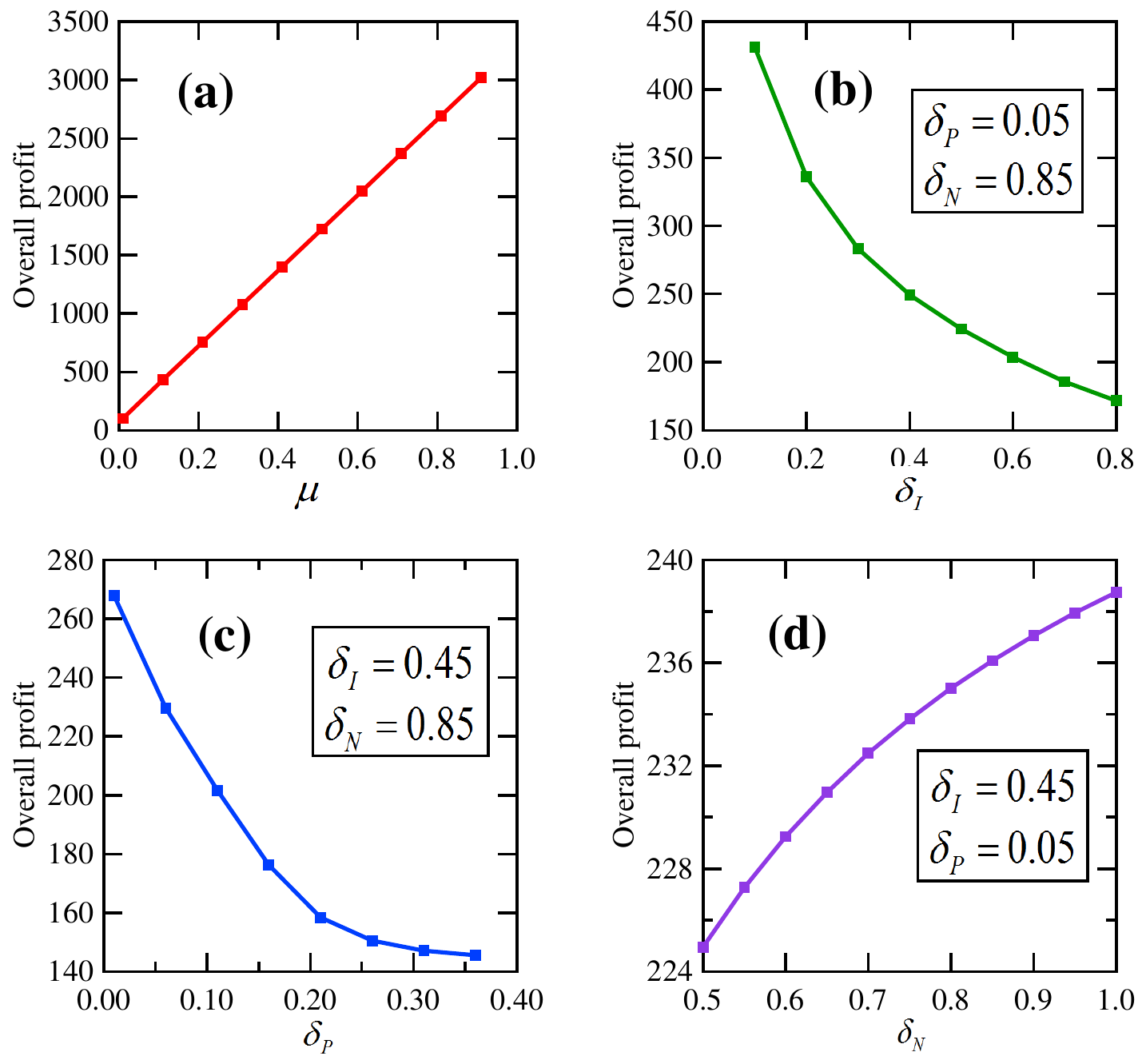}
	\caption{The expected overall profit versus (a) the entrance rate, (b) the I-exit rate, (c) the P-exit rate, and (d) the N-exit rate.}
\end{figure}

The entrance rate can be enhanced by launching a viral marketing (VM) campaign, that is, marketers develop a marketing message and encourage customers to forward this message to their contacts. There are quite a number of successful VM cases: Hotmail generated 12 million subscribers in just 18 months with a marketing budget of only \$50,000, and Unilever’s Dove Evolution campaign generated over 2.3 million views in its first 10 days. These VM campaigns were successful in part because the marketers effectively utilized VM's unique potential to reach large numbers of potential customers in a short period of time at a lower cost.

The I-exit rate or the P-exit rate can be reduced by enhancing customers' experiences with the purchased products and taking promotional measures such as discounting and distributing coupons.

Typically, the N-exit rate is uncontrollable.

\subsection{The comment rates}

Comprehensive simulation experiments show that, typically, the influence of the two comment rates on the expected overall profit is as shown in Fig. 13. In general, the following conclusions are drawn.

\begin{enumerate}
	\item[(a)] With the rise of the P-comment rate, the overall profit goes up.
	\item[(b)] With the rise of the N-comment rate rate, the overall profit goes down.
\end{enumerate}

\begin{figure}[ht]
	\setlength{\abovecaptionskip}{0.cm}
	\setlength{\belowcaptionskip}{-0.cm}
	\centering
	\includegraphics[width=0.7\textwidth]{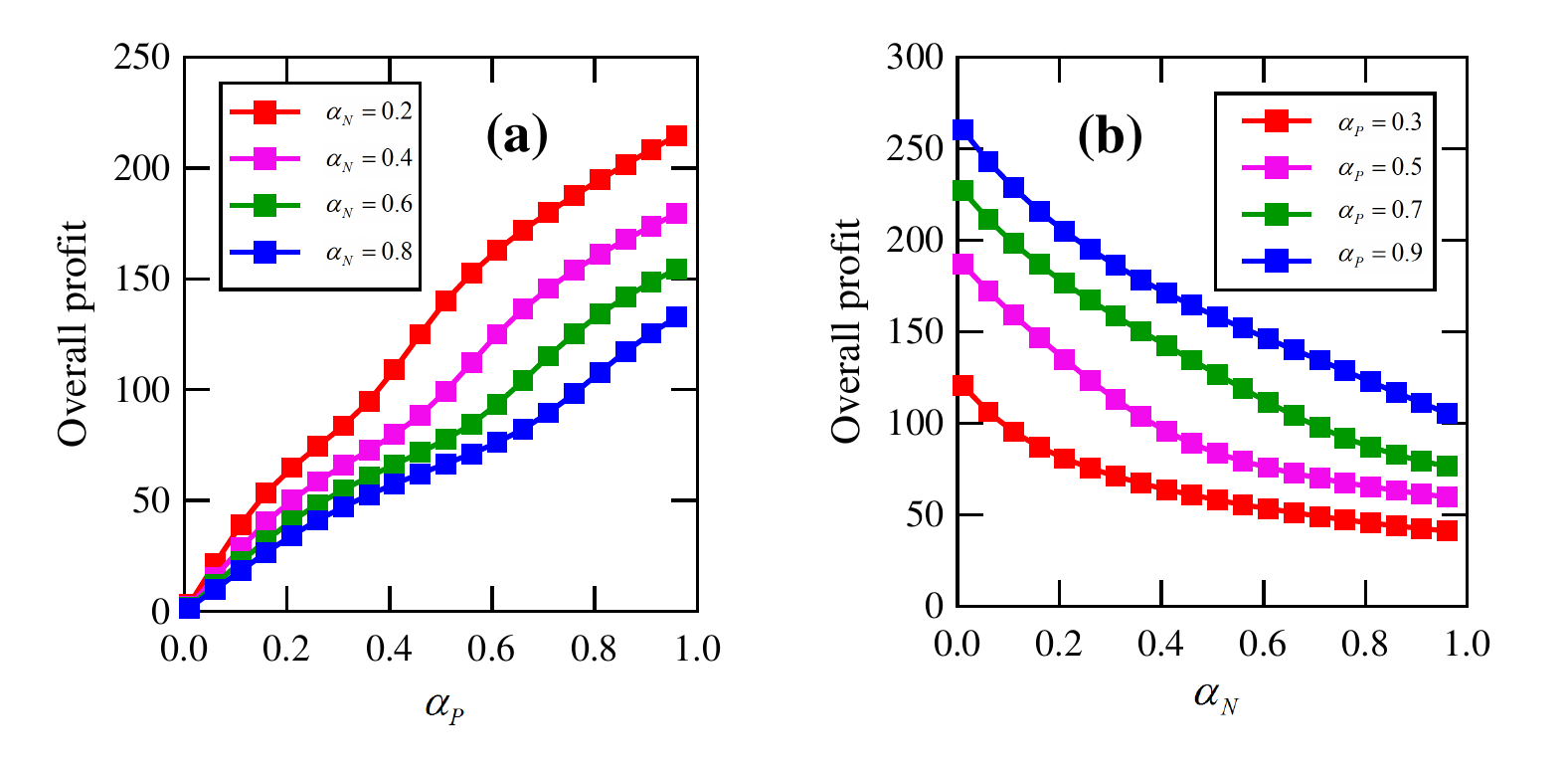}
	\caption{The expected overall profit versus (a) the P-comment rate, and (b) the N-comment rate.}
\end{figure}

The P-comment rate can be enhanced by  enhancing customers' experiences. Also, the N-comment rate can be enhanced by  enhancing customers' experiences.

\subsection{The infection forces}

Comprehensive simulation experiments show that, typically, the influence of the two infection forces on the expected overall profit is as shown in Fig. 14. In general, the following conclusions are drawn.

\begin{enumerate}
	\item[(a)] With the rise of the P-infection force, the overall profit goes up.
	\item[(b)] With the rise of the N-infection force, the overall profit goes down.
\end{enumerate}

\begin{figure}[ht]
	\setlength{\abovecaptionskip}{0.cm}
	\setlength{\belowcaptionskip}{-0.cm}
	\centering
	\includegraphics[width=0.7\textwidth]{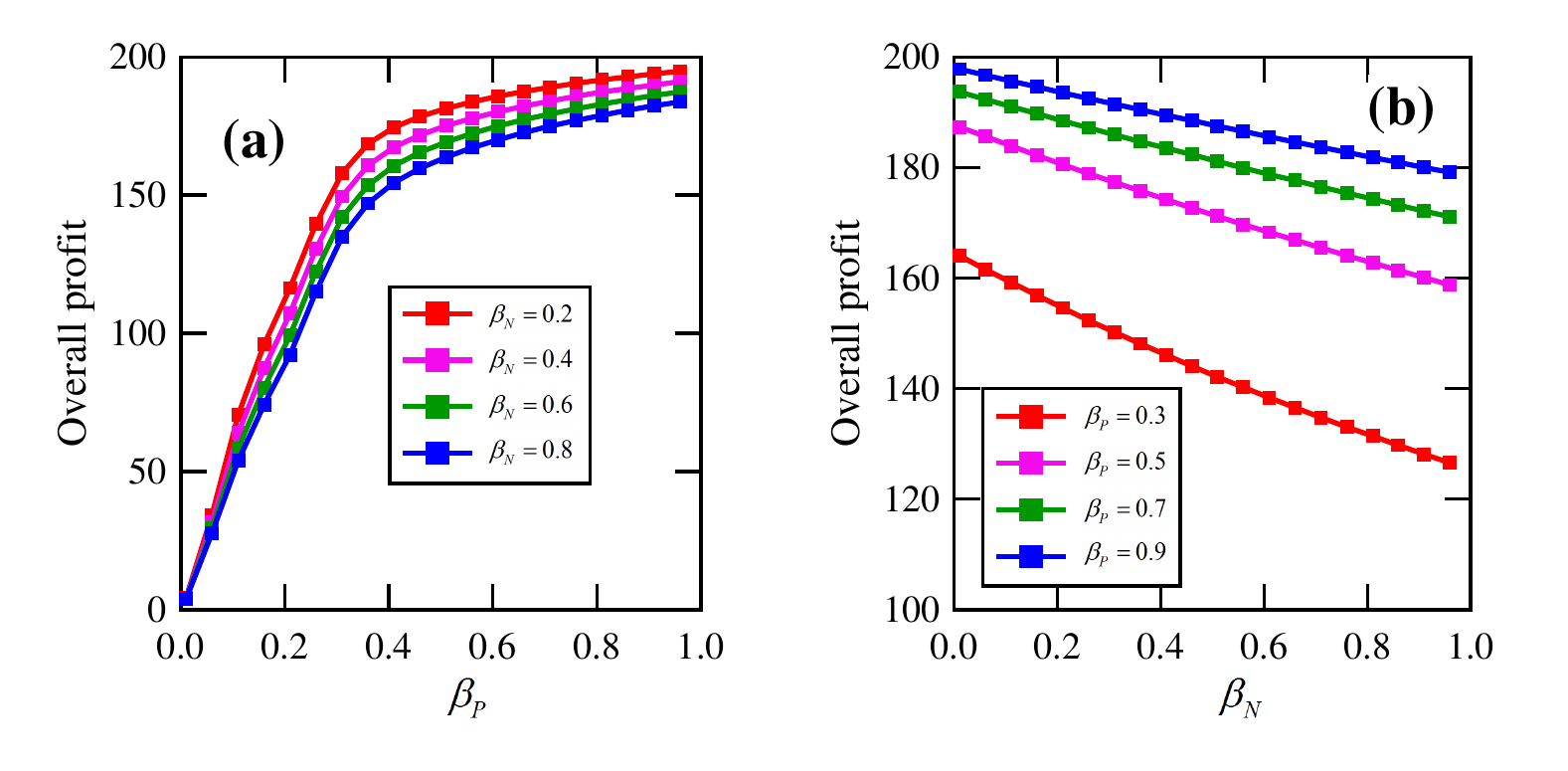}
	\caption{The expected overall profit versus (a) the P-infection force, and (b) the N-infection force.}
\end{figure}

The P-infection force can be enhanced by enhancing customers' experiences and, hence, earning positive WOM. Also, the N-infection force can be reduced by enhancing customers' experiences.

\subsection{The viscosity rates}

Comprehensive simulation experiments show that, typically, the influence of the two viscosity rates on the expected overall profit is as shown in Fig. 15. In general, the following conclusions are drawn.

\begin{enumerate}
	\item[(a)] With the rise of the I-viscosity rate, the overall profit goes up.
	\item[(b)] There is a threshold $\theta$ such that (1) when the P-viscosity is lower than $\theta$, the expected overall profit goes up with the rise of the P-viscosity, and (2) when the P-viscosity exceeds $\theta$, the expected overall profit goes down with the rise of the P-viscosity.
\end{enumerate}

\begin{figure}[ht]
	\setlength{\abovecaptionskip}{0.cm}
	\setlength{\belowcaptionskip}{-0.cm}
	\centering
	\includegraphics[width=0.8\textwidth]{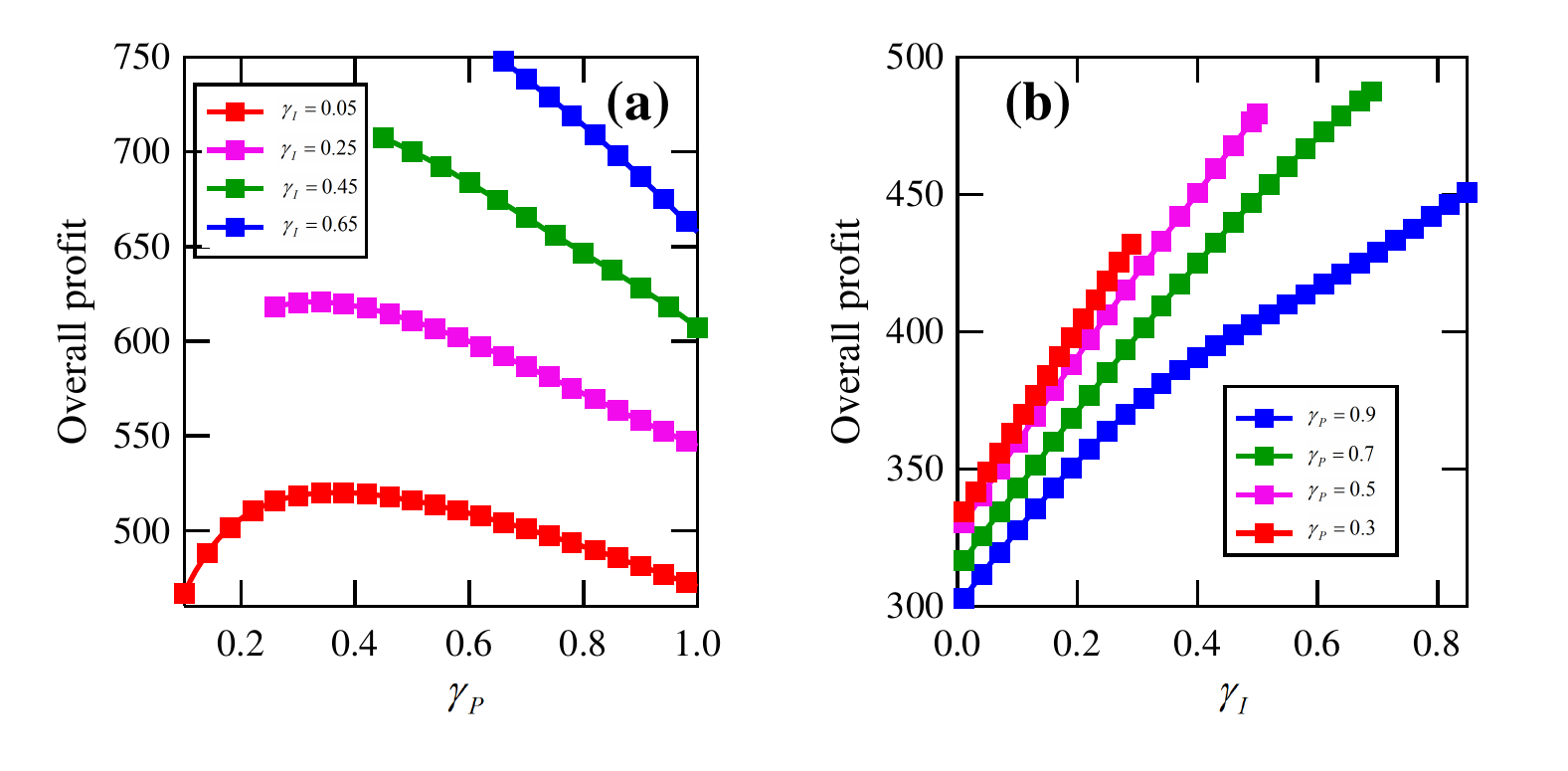}
	\caption{The overall profit versus (a) the P-viscosity rate, and (b) the I-viscosity rate.}
\end{figure}

The I-viscosity rate can be enhanced by enhancing customers' experiences or by taking promotional measures.

\section{Conclusions and remarks}

The issue of maximizing the overall profit of a WOM-based promotion campaign has been modeled as a constrained optimization problem. Through comprehensive computer simulations, the influence of different factors on the profit has been revealed. On this basis, some promotion strategies have been suggested.

Toward this direction, great efforts have yet to be made. It is well known that the structure of the underlying WOM network has significant influence on the performance of a viral marketing \cite{Bampo2008, Siri2013}. The proposed SIPNS model is a population-level model and hence does not allow the analysis of this influence. To reveal the impact of the WOM network on the marketing profit, a network-level epidemic model \cite{Satorras2001, Satorras2002, YangLX2014a, YangLX2017a} or an individual-level epidemic model \cite{Mieghem2009, Sahneh2013, YangLX2015a, YangLX2017b, YangLX2017c} is an appropriate choice. The profit model presented in this paper builds on the uniform-profit assumption. However, in everyday life different products may have seperate profits. Hence, it is of practical importance to construct a non-uniform profit model. Also, a customer may purchase more than one items a time, and the corresponding model is yet to be developed. Typically, a viral marketing campaign is subject to a limited budget. The dynamic optimal control strategy against malicious epidemics \cite{Preciado2014, Shakeri2015, YangLX2016, Nowzari2016, ZhangTR2017} may be borrowed to the analysis of viral marketing so as to achieve the maximum possible net profit.

\section*{Acknowledgments}

This work is supported by Natural Science Foundation of China (Grant Nos. 61572006, 71301177), National Sci-Tech Support Program of China (Grant No. 2015BAF05B03), Basic and Advanced Research Program of
Chongqing (Grant No. cstc2013jcyjA1658), and Fundamental Research Funds for the Central Universities (Grant No. 106112014CDJZR008823).




\bibliography{<your-bib-database>}

\begin{thebibliography}{43}


\bibitem[1]{Armstrong2012}
G. Armstrong, P. Kotler, Marketing: An introduction (11th Edition), Prentice Hall, 2012.

\bibitem[2]{Grewal2016}
D. Grewal, M. Levy, Marketing (5th Edition), McGraw-Hill Education, 2016.

\bibitem[3]{Goldenberg2001}
J. Goldenberg, B. Libai, E. Muller, Talk of the network: A complex systems look at the underlying process of word-of-mouth, Marketing Letters, 2001, 12(3): 211-223.

\bibitem[4]{Mahajan1984}
V. Mahajan, E. Muller, R.A. Kerin, Introduction strategy for new products with positive and negative word-of-mouth, Management Science, 1984, 30(12): 1389-1404.

\bibitem[5]{Anderson1998}
E. Anderson, Customer satisfaction and word-of-mouth, Journal of Service Research, 1998, 1(1): 5-17.

\bibitem[6]{Herr1991}
P.M. Herr, F.R. Kardes, J. Kim, Effects of word-of-mouth and product attribute information on persuasion: An accessibility-diagnosticity perspective, Journal of Consumer Research, 1991, 17, 454-462.

\bibitem[7]{Charlett1995}
D. Charlett, R. Garland, N. Marr, How damaging is negative word of mouth? Marketing Bulletin, 1995, 6: 42-50.

\bibitem[8]{Sweeney2007}
J.C. Sweeney, G.N. Soutar, T. Mazzarol, The differences between positive and negative word-of-mouth: Emotion as a differentiator? S. Gopalan, N. Taher (eds), Viral Marketing: Concepts and Cases, Icfai University Press, 2007, pp. 156-168.

\bibitem[9]{Ahmad2014}
N. Ahmad, J. Vveinhardt, R.R. Ahmed, Impact of word of mouth on consumer buying decision, European Journal of Business and Management, 2014, 6(31): 394-403.

\bibitem[10]{Misner1999}
I.R. Misner, The World's Best Known Marketing Secret: Building Your Business with Word-of-Mouth Marketing, 2nd ed. Bard Press, 1999.

\bibitem[11]{Chevalier2006}
J. Chevalier, M. Dina, The effect of word of mouth on sales: online book reviews, Journal of Marketing Research, 2006, 43: 345-354.

\bibitem[12]{Trusov2009}
M. Trusov, R.E. Bucklin, K. Pauwels, Effects of word-of-mouth versus traditional marketing: Findings from an Internet social networking sites, Journal of Marketing, 2009, 73(5): 90-102.

\bibitem[13]{Peres2010}
R. Peres, E. Muller, V. Mahajan, Innovation diffusion and new product growth models: A critical review and research directions, International Journal of Research in Marketing, 2010,  27: 91-106.

\bibitem[14]{Kempe2005}
D. Kempe, J. Kleinberg, E, Tardos, Influential nodes in a diffusion model for social networks, Proceedings of the 32nd International Conference on Automata, Languages and Programming, 2005, 1127-1138.

\bibitem[15]{ChenW2009}
W. Chen, Y. Wang, S. Yang, Efficient influence maximization in social networks, Proceedings of the 15th ACM SIGKDD International Conference on Knowledge Discovery and Data Mining, 2009, pp. 199-208.

\bibitem[16]{Hinz2011}
O. Hinz, B. Skiera, C. Barrot, J.U. Becker, Seeding strategies for viral marketing: An empirical comparison, Journal of Marketing, 2011, 75: 55-71. 

\bibitem[17]{Dinh2014}
T.N. Dinh, H. Zhang, D.T. Nguyen, M.T. Thai, Cost-effective viral marketing for tme-critical campaigns in large-scale social networks, IEEE/ACM Transactions on Networking, 2014, 22(6): 2001-2011.

\bibitem[18]{Mochalova2014}
A. Mochalova, A. Nanopoulos, A targeted approach to viral marketing, Electronic Commerce Research and Applications, 2014, 13: 283-294.

\bibitem[19]{Kempe2015}
D. Kempe, J. Kleinberg, E, Tardos, Maximizing the spread of influence through a social network, THEORY OF COMPUTING, 2015, 11(4): 105-147.

\bibitem[20]{Samadi2016}
M. Samadi, A. Nikolaev, R. Nagi, A subjective evidence model for influence maximization in social networks, Omega, 2016, 59(B): 263-278.

\bibitem[21]{ZhangHY2016a}
H. Zhang, D.T. Nguyen, S. Das, H. Zhang, M.T. Thai, Least cost influence maximization across multiple social networks, IEEE/ACM Transactions on Networking, 2016, 24(2): 929-939.

\bibitem[22]{Bharathi2007}
S. Bharathi, D. Kempe, M. Salek, Competitive influence maximization in social networks, X. Deng, F. C. Graham (eds) Internet and Network Economics, WINE2007, Lecture Notes in Computer Science, vol. 4858, 2007. 

\bibitem[23]{ZhangHY2016b}
H. Zhang, H. Zhang, A. Kuhnle, M.T. Thai, Profit maximization for multiple products in online social networks, Proceedings of the IEEE International Conference on Computer Communications (INFOCOM), 2016.

\bibitem[24]{Bass1969}
F.M. Bass, A new product growth for model consumer durables, Management Science, 1969, 15(5): 215-227.

\bibitem[25]{Gardner2013}
J.T. Gardner, K. Sohn, J.Y. Seo, J.L. Weaver, Analysis of an epidemiological model of viral marketing: when viral marketing efforts fall flat, Journal of Marketing Development and Competitiveness, 2013, 7(4): 25-46.

\bibitem[26]{Sohn2013}
K. Sohn, J. Gardner, J. Weaver, Viral marketing --- more than a buzzword, Journal of Applied Business and Economics, 2013, 14(1): 21-42.

\bibitem[27]{LiS2013}
S. Li, Z. Jin, Global dynamics analysis of homogeneous new products
diffusion model, Discrete Dynamics in Nature and Society, 2013, 2013: 158901.

\bibitem[28]{LiS2014}
S. Li, Z. Jin, Modeling and analysis of new products diffusion on
heterogeneous networks, Journal of Applied Mathematics, 2014, 2014: 940623.

\bibitem[29]{Rodrigues2015}
H.S. Rodrigues, M. Fonseca, Viral marketing as epidemiological model, Proceedings of the 15th International Conference on Computational and Mathematical Methods in Science and Engineering, 2015, pp. 946-955. 

\bibitem[30]{Rodrigues2016}
H.S. Rodrigues, M. Fonseca, Can information be spread as
a virus? Viral marketing as epidemiological model, Mathematical Methods in the Applied Sciences, 2016, 39: 4780-4786.

\bibitem[31]{JiangP2017}
P. Jiang, X. Yan, L. Wang, A viral product diffusion model to forecast the market performance of products, Discrete Dynamics in Nature and Society, 2017, 2017: 9121032.

\bibitem[32]{WangWD2003}
W. Wang, P. Fergola, C. Tenneriello, Innovation diffusion model in
patch environment, Applied Mathematics and Computation, 2003, 134: 51-67.

\bibitem[33]{YuY2003}
Y. Yu, W. Wang, Y. Zhang, An innovation diffusion model for three competitive products, Computers and Mathematics with Applications, 2003, 46: 1473-1481.

\bibitem[34]{WeiXT2013}
X. Wei, N. Valler, B.A. Prakash, I. Neamtiu, M. Faloutsos, C. Faloutsos, Competing memes propagation on networks: A network science perspective, IEEE Journal on Selected Areas in Communications, 2013, 31(6): 1049-1060. 

\bibitem[35]{Bampo2008}
M. Bampo, M. Ewing, D. Mather, D. Stewart, M. Wallace, The effects of the social structure of digital networks on viral marketing performance, Information Systems Research, 2008, 19(3): 273-290.

\bibitem[36]{Siri2013}
A. Siri, T. Thaiupathump, Measuring the performance of viral marketing based on the dynamic behavior of social Networks, Proceedings of the IEEE International Conference on Industrial Engineering and Engineering Management, 2013, 432.

\bibitem[37]{Satorras2001}
P. Pastor-Satorras, A. Vespignani, Epidemic spreading in scale-free networks, Physical Review Letters, 2001, 86(14): 3200.

\bibitem[38]{Satorras2002}
P. Pastor-Satorras, A. Vespignani, Epidemic dynamics in finite size scale-free networks, Physical Review E, 2002, 65(3): 035108.

\bibitem[39]{YangLX2014a}
L.X. Yang, X. Yang, The spread of computer viruses over a reduced scale-free network, Physica A, 2014, 396(15): 173-184.

\bibitem[40]{YangLX2017a}
L.X. Yang, X. Yang, The effect of network topology on the spread of computer viruses: a modelling study, International Journal of Computer Mathematics, DOI: 10.1080/00207160.2016.1226499.

\bibitem[41]{Mieghem2009}
P. Van Mieghem, J. Omic, R. Kooij, Virus spread in networks, IEEE/ACM Transactions on Networking, 2009, 17(1): 1-14.

\bibitem[42]{Sahneh2013}
F.D. Sahneh, C. Scoglio, P. Van Mieghem, Generalized epidemic mean-field model for spreading processes over multilayer complex networks, IEEE/ACM Transactions on Networking, 2013, 21(5): 1609-1620.

\bibitem[43]{YangLX2015a}
L.X. Yang, M. Draief, X. Yang, The impact of the network topology on the viral prevalence: a node-based approach, Plos One, 2015, 10: e0134507.

\bibitem[44]{YangLX2017b}
L.X. Yang, M. Draief, X. Yang, Heterogeneous virus propagation in networks: a theoretical study. Mathematical Methods in the Applied Sciences, 2017, 40(5): 1396-1413.

\bibitem[45]{YangLX2017c}
L.X. Yang, X. Yang, Y. Wu, The impact of patch forwarding on the prevalence of computer virus: A theoretical assessment approach, Applied Mathematical Modelling, 2017, 43: 110-125.

\bibitem[46]{Preciado2014}
V.M. Preciado, M. Zargham, C. Enyioha, A. Jadbabaie, G. Pappas, Optimal resource allocation for network protection against spreading processes, IEEE Transactions on Control of Network Systems, 2014, 1(1): 99-108.

\bibitem[47]{Shakeri2015}
H. Shakeri, F.D. Sahneh, C. Scoglio, Optimal information dissemination strategy to promote preventive behaviours in multilayer networks. Mathematical Bioscience and Engineering, 2015, 12(3): 609-623.

\bibitem[48]{YangLX2016}
L.X. Yang, M. Draief, X. Yang, The optimal dynamic immunization under a controlled heterogeneous node-based SIRS model, Physic A, 2016, 450: 403-415.

\bibitem[49]{Nowzari2016}
C. Nowzari, V.M. Preciado, G.J. Pappas, Analysis and control of epidemics: A survey of spreading processes on complex networks, IEEE Control Systems, 2016, 36(1): 26-46.

\bibitem[50]{ZhangTR2017}
T.R. Zhang, L.X. Yang, X. Yang, Y. Wu, Y.Y. Tang, The dynamic malware containment under a node-level epidemic model with the alert compartment, Physica A, 2017, 470: 249-260.

\bibitem[51]{BiJC2017}
J. Bi, X. Yang, Y. Wu, Q. Xiong, J. Wen, Y.Y. Tang, On the optimal dynamic control strategy of disruptive computer virus, Discrete Dynamics in Nature and Society, 2017, 2017: 8390784.

\end{thebibliography}



\end{document}